\documentclass[ reprint, superscriptaddress, floatfix, amsmath,amssymb, aps, prx ]{revtex4-2}

\usepackage[utf8]{inputenc}
\usepackage{amsmath,amssymb}
\usepackage{wrapfig}
\usepackage{graphicx}
\usepackage{bm}
\usepackage[colorlinks=true,linkcolor=blue,citecolor=blue,urlcolor=blue]{hyperref}
\usepackage{makecell}
\usepackage{siunitx}
\usepackage{cases}

\newcommand{\citeasnoun}[1]{Ref.~\cite{#1}}

\newcommand{\Figref}[1]{Figure~\ref{fig:#1}}
\newcommand{\figref}[1]{Fig.~\ref{fig:#1}}
\renewcommand{\eqref}[1]{Eq.~(\ref{eq:#1})}

\begin{document}

\preprint{APS/123-QED}

\title{High-NA optical edge detection via optimized multilayer films}

\author{Wenjin Xue}
\affiliation{Department of Electrical Engineering and Energy Sciences Institute, Yale University, New Haven, Connecticut 06511, USA}
\author{Owen D. Miller}
\affiliation{Department of Applied Physics and Energy Sciences Institute, Yale University, New Haven, Connecticut 06511, USA}

\date{\today}

\begin{abstract}
    There has been a significant effort to design nanophotonic structures that process images at the speed of light. A prototypical example is in edge detection, where photonic-crystal-, metasurface-, and plasmon-based designs have been proposed and in some cases experimentally demonstrated. In this work, we show that multilayer optical interference coatings can achieve visible-frequency edge detection in transmission with high numerical aperture, two-dimensional image formation, and straightforward fabrication techniques, unique among all nanophotonic approaches. We show that the conventional Laplacian-based transmission spectrum may not be ideal once the scattering physics of real designs is considered, and show that better performance can be attained with alternative spatial filter functions. Our designs, comprising alternating layers of Si and SiO$_2$ with total thicknesses of only $\approx \SI{1}{\um}$, demonstrate the possibility for optimized multilayer films to achieve state-of-the-art edge detection, and, more broadly, analog optical implementations of linear operators.
\end{abstract}

\maketitle
\section{Introduction}
In this article, we show that optimally designed multilayer dielectric films can achieve high-numerical-aperture optical edge detection. In a typical scenario, one might have a scene or object illuminated by a laser or narrow-bandwidth source, in which case an optical device that generates an image of edges offers the prospect for speed-of-light detection~\cite{solli2015analog,zhu2017plasmonic,guo2018photonic,kwon2018nonlocal,zhu2019generalized,zhou2019optical,wesemann2019selective,cordaro2019high,zhou2020flat,kwon2020dual,zhu2021topological}. Of the many edge-detection designs to date~\cite{zhu2017plasmonic,guo2018photonic,kwon2018nonlocal,zhu2019generalized,zhou2019optical,wesemann2019selective,cordaro2019high,zhou2020flat,kwon2020dual,zhu2021topological}, none offers all three of: high-numerical-aperture, two-dimensional image formation, the prospect for realistic fabrication, and transmission-mode operation. Multilayer films are commonly fabricated to high precision for wide-ranging applications~\cite{yeh1988optical,faist1994quantum}, and guarantee a two-dimensional field of view by their rotational symmetry. Thus the key question is whether they can achieve high-fidelity, transmission-mode-based edge detection for a large numerical aperture, which we show is indeed possible with structural optimization. The canonical approach to achieving edge-detection behavior~\cite{silva2014performing,lissberger1970optical} is to target a transmission profile that scales with $k_{\rho}^2$, for in-plane wavevector $k_{\rho}$, as such a profile will mimic the effect of the in-plane Laplacian operator, $\nabla_{\perp}^2$, on the incoming field. We show that targeting such a quadratic profile with a multilayer structure can successfully produce an effective design. Yet, the design is imperfect, and the requirement of a $k_{\rho}^2$ profile is an over-prescription of the response function. \emph{Any} transmission response that acts as a high-pass filter, i.e., which filters out small spatial frequencies (corresponding to nearly constant in-plane spatial modes), can produce high-quality edges. To demonstrate this, we show that a transmission profile that scales as the cube of the incident angle, $\sim \theta^3$, can produce even higher-quality images for the same design parameters. Our designs offer the highest theoretical performance to date, should be straightforward to fabricate, and reveal the potential of such multilayer structures for analog optical devices.

\begin{figure}[!tbh]
    \includegraphics[width=\linewidth]{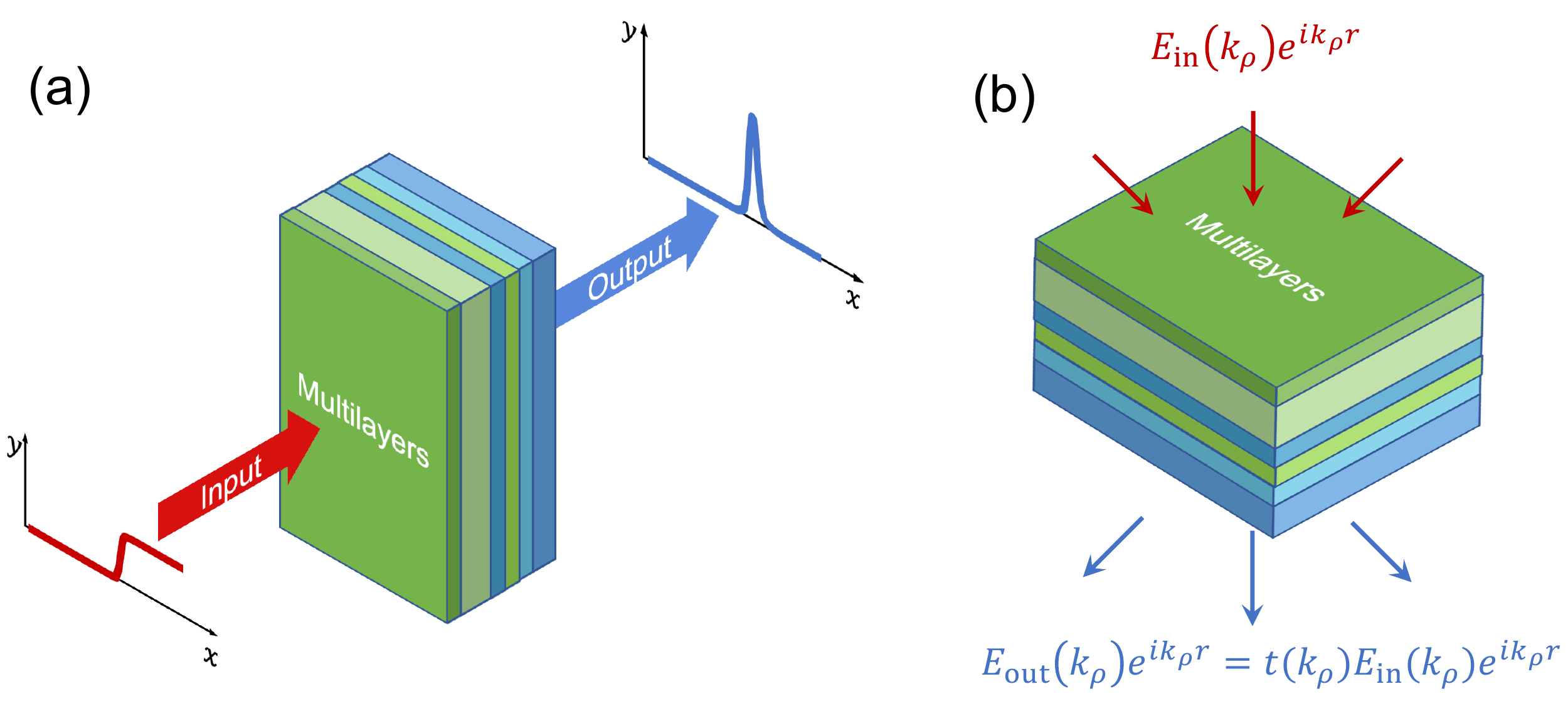}
    \centering
    \caption{(a) Illustration of edge detection with multilayer films. (b) Edge detection can be achieved by engineering the transmission spectrum in Fourier space. For a coherent incident wave decomposed into its plane-wave constituents, labeled by in-plane wavevector $k_{\rho}$, the transmission coefficient $t(k_{\rho})$ determines the image at the output plane.}
    \label{fig:figure1}
\end{figure}

The essential feature of optical analog edge detection, as shown in \figref{figure1}, is to engineer the light field of a coherent image, isolating its edges in the transmission or reflection spectrum of an optical device. A classic approach~\cite{goodman2005introduction} to analog edge detection is to use a lens to Fourier transform the incoming waves and an aperture to filter out the low in-plane wavevector components, with two free-space propagation regions to allow the evolution of the wave field to achieve the Fourier and Inverse-Fourier Transforms. The key drawback is that the setup must be large and bulky to accommodate the free-space propagation. An emerging alternative is to use coherent scattering effects to isolate edges in a more compact device architecture, including photonic-crystal slabs~\cite{guo2018photonic,zhou2020flat,kwon2020dual}, dielectric metasurfaces~\cite{zhou2019optical,zhu2019generalized,cordaro2019high}, plasmonic films~\cite{zhu2017plasmonic,wesemann2019selective}, dielectric interface~\cite{zhu2021topological} and split-ring-resonator metamaterials~\cite{kwon2018nonlocal}. The incoming field, for a given polarization at a frequency of interest, can be written as a linear combination of plane waves, $\int E_{\rm in}(k_{\rho}) e^{i(k_{\rho}\rho+k_z z)} \,{\rm d}k_{\rho}$. One method to identify edges is to try to identify a structure whose wavevector-dependent transmission or reflection mimics the in-plane Laplacian operator, $\nabla_{\perp}^2$. In  the spatial Fourier space, the Laplacian corresponds to multiplying the incoming plane waves by $k_{\rho}^2$, suppressing low spatial frequencies relative to high ones. The effectiveness of the Laplacian can be attributed to the fact that edges are high-spatial-frequency components of images, whereas a low-contrast background comprises primarily low spatial frequencies. Although the devices discussed above~\cite{zhu2017plasmonic,guo2018photonic,kwon2018nonlocal,zhu2019generalized,zhou2019optical,wesemann2019selective,cordaro2019high,zhou2020flat,kwon2020dual,zhu2021topological} have shown the possibility for discriminating edges within an image, there have been key drawbacks to each approach: they have been limited to one-dimensional (or partial-two-dimensional) image formation, non-ideal reflection-mode operation, small numerical aperture, and/or hard-to-fabricate structures.

\section{Optimization method}
Multilayer films (i.e., optical interference coatings) have well-established fabrication techniques~\cite{jaeger2002introduction,seshan2001handbook}, and their optical response is necessarily isotropic under rotations around their propagation axis. Thus if their transmission coefficients can be optimized to have the right profile, they can simultaneously satisfy the three key requirements (high-NA 2D field of view, simple fabrication, and transmission-mode operation). We define a target transmission coefficient, $t_{\rm target}$, as a function of angle (or, equivalently, in-plane wavevector), which serves as the ideal transmission function for edge detection. We take the allowed materials in the multilayer to be given, and use the \emph{thicknesses} of the corresponding layers, $w_{\ell}$ for each layer $\ell$, to be the designable degrees of freedom. Our optimization problem, then, is to minimize the error between the designed and targeted transmission coefficients over a sufficiently dense discrete set of incoming angles $\theta$:
\begin{align}
    \min_{w_{\ell}} \sum_{\theta} \left|t(\theta;w_{\ell}) - t_{\rm target}(\theta)\right|^2
    \label{eq:opt_prob}
\end{align}
where $t(\theta;w_{\ell})$ is the transmission coefficient of a given multilayer stack. One can optimize the transmission coefficient for edge-image formation at any output plane beyond the multilayer; to demonstrate how compact this approach can be, we take the image plane to be the exterior of the rear surface itself.

For optimization of a large number of layers, one must be able to rapidly compute gradients of the objective function, \eqref{opt_prob}, with respect to the many degrees of freedom. Here we briefly outline how the gradients are computed, using a method similar to that of the ``needle'' approach to multilayer-film design~\cite{vd1978fast,peng1985derivatives,tikhonravov1982synthesis,tikhonravov1996application}. In a multilayer medium, the continuous translational and rotational symmetry prevents coupling between different in-plane wavevectors. At each wavevector, the standard matrix approach~\cite{yeh1988optical} connects the forward- and backward-going wave amplitudes in the incident region to the equivalent amplitudes in the transmission region through matrices $P_\ell$ and $D_\ell$ and that represent propagation through, and interface reflections at, layer $\ell$. The reflection coefficient $r$ and transmission coefficient $t$ satisfy the matrix equation~\cite{yeh1988optical}:
\begin{align}
                        \begin{pmatrix}
                            1 \\
                            r
                        \end{pmatrix} &= 
                        D_0^{-1}\left[ \displaystyle\prod_{\ell=1}^{N} D_\ell P_\ell D_\ell^{-1}\right]D_s
                        \begin{pmatrix}
                            t \\
                            0
                        \end{pmatrix}\nonumber\\
                        &=
                        \begin{pmatrix}
                            M_{11}  & M_{12} \\
                            M_{21} & M_{22}
                        \end{pmatrix}  
                        \begin{pmatrix}
                            t \\
                            0
                        \end{pmatrix}.
                        \label{eq:TM}
\end{align}
where the ``$1$'' on the left-hand side represents the incident-wave normalization, and the reflection coefficient, transmission coefficient, and the matrices $P_{\ell}$ and $D_{\ell}$ all vary with wavevector $k_{\rho}$. One can then solve for $r$ and $t$ from the components of the $M$ matrix; in particular, $t$ is given by $t = 1/M_{11}$ (\citeasnoun{yeh1988optical}). The derivative of $t$ with respect to the thickness of region $i$ is then given by ${\rm d}t/{\rm d}w_i = -(1/M_{11}^2) {\rm d} M_{11} / {\rm d} w_i$. A change in the thickness of layer $i$, while keeping all other layer thicknesses fixed, does not affect interface transmission and reflection, and will only incur changes in the propagation matrix $P_i$ in the product of \eqref{TM}. The derivative of the $M$ matrix can then be written:
\begin{align}
    &\frac{{\rm d}M}{{\rm d} w_i} =\nonumber\\
    &\quad D_0^{-1} \left[ \displaystyle\prod_{\ell=1}^{i-1} D_\ell P_\ell D_\ell^{-1}\right] D_i
    \frac{{\rm d} P_i}{{\rm d} w_i}D_i^{-1} \left[ \displaystyle\prod_{\ell=i+1}^{N}D_\ell P_\ell D_\ell^{-1}\right] D_s
    \label{eq:dMdx} 
\end{align}
which can be rapidly computed.

\begin{figure*}[!htb]
    \includegraphics[width=\textwidth]{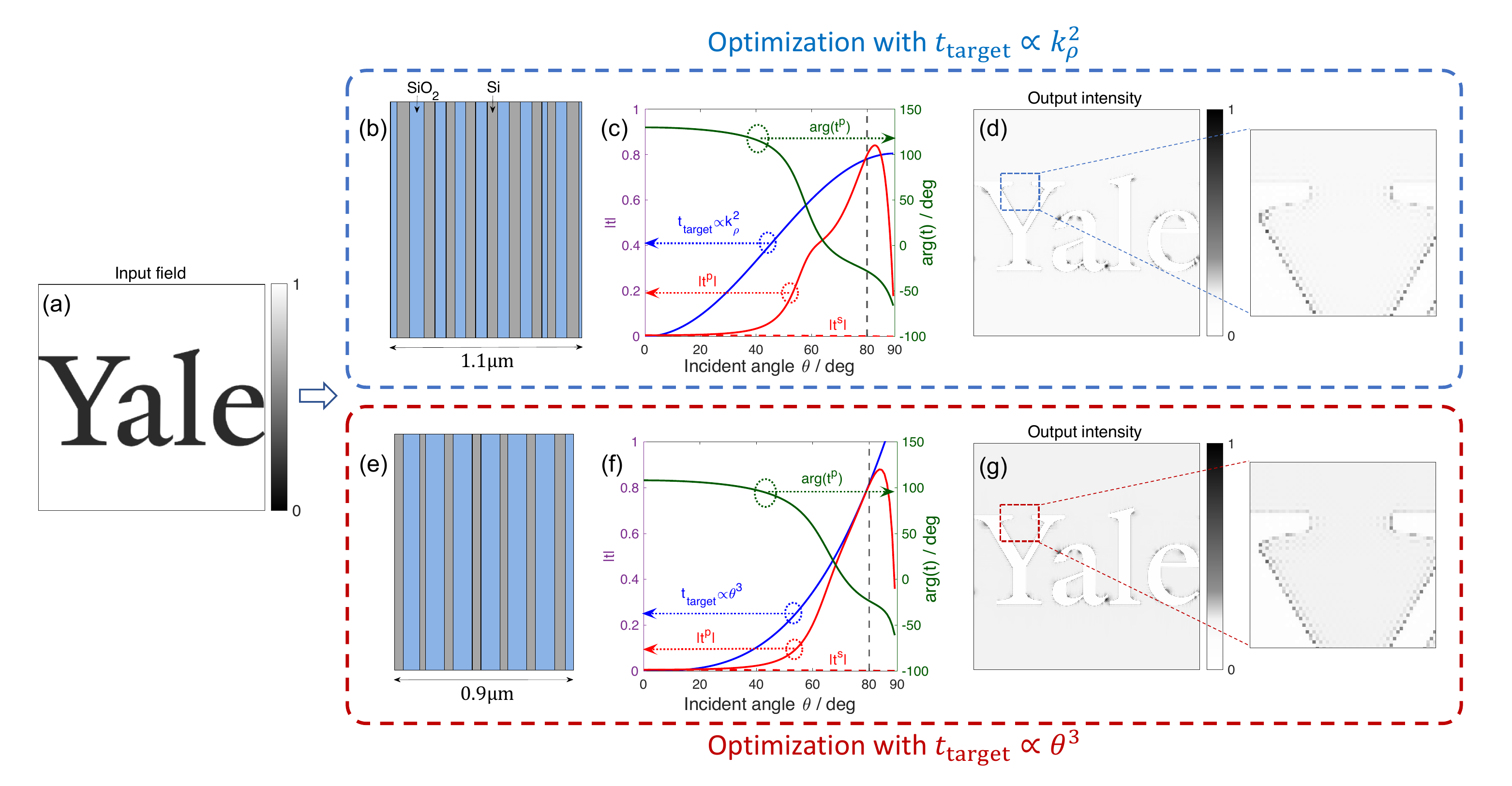}
    \centering
    \caption{Multilayer edge-detection designs for operating wavelength of \SI{700}{nm}. (a) Input image, with edges oriented in every direction. (b) Optimal design for a target transmission coefficient proportional to the square of the in-plane wavevector. The design comprises alternating layers of Si and $\textrm{SiO}_2$. (c) Transmission coefficients, in magnitude and phase, of the target and optimal designs, showing close but imperfect correspondence. (The target design has phase 0 at all $\theta$.) At glancing incidence, $\theta \rightarrow \ang{90}$, any multilayer film must have zero transmission, but the multilayer design is able to achieve increasing transmission up to \ang{80}, corresponding to an NA of 0.98. (d) Output image, clearly showing edges with minimal background noise or interference. (e-g) Same plots as (b-d) but for a target transmission coefficient proportional to the cube of the incident angle, demonstrating that other spatial filter functions can demonstrate as good or better performance as the quadratic Laplacian spectrum.}
    \label{fig:figure2}
\end{figure*}

\section{Results}
\Figref{figure2} demonstrates the capability for high-efficacy edge detection with a computationally optimized multilayer stack. We consider up to 20 alternating layers of Si and SiO$_2$, with refractive indices $3.77+0.01i$ and $1.47$, respectively, for light incident at \SI{700}{nm} wavelength. We choose materials such as silicon for their large refractive indices, which facilitates stronger interactions with light~\cite{shim2020optical} and high-performance designs. Although one might be concerned about the lossiness of silicon, the designs are ultra-thin, with total silicon thicknesses on the order of half a micron, well below the $>\SI{5}{\um}$ absorption depth of silicon at \SI{700}{nm} wavelength. To perform the optimization, we use the gradient computed via \eqref{dMdx} in a local, gradient-based interior-point method~\cite{byrd2000trust,byrd1999interior,waltz2006interior,MATLAB} that is run until convergence. Simulations of image formation are done by Fourier transforming the incident field, multiplying by the wavevector-dependent transmission, and then inverse Fourier transforming at the output plane. As our input we consider a Yale logo, containing edges oriented in almost every direction. We take the image to occupy a numerical aperture of $\approx 0.98$, corresponding to polar angles ranging from \ang{0} to \ang{80}.

In one set of optimizations, we target a Laplacian-based transmission coefficient function given by $t_{\rm target}(\theta) = \alpha k_{\rho}^2$, where $\alpha$ is a constant optimization hyperparameter. Ideal design would have both polarizations follow the lineshape of $t_{\rm target}$; in practice, however, we find that such designs appear to be impossible to achieve in the multilayer form factor. An alternative solution, halving the brightness, is to fully reflect one polarization and perform the filtering operation in the transmission spectrum of the other polarization. (Note that the orientations of the usual $s$ and $p$ polarizations of plane waves are wavevector-dependent, and a linearly polarized incoming wave contains each.) After running the optimization for many values of $\alpha$ and many layer-thickness initializations, the design shown in \figref{figure2}(b), slightly more than \SI{1}{\um} in total thickness, emerges as optimal (cf. SM for detailed design data). \Figref{figure2}(c) shows the actual transmission coefficient of the design (red and green lines), compared to the target (blue). The transmission of $s$-polarized waves is nearly 0 for all incident angles (red dashed line), so only $p$-polarized waves (red solid line) contribute to the final edge detection. The targeted quadratic $k_{\rho}^2$ distribution is not exactly mimicked, but the variation in phase is less than $\pi$ over the whole angular range, which ensures effective interference at the image plane. Interestingly, the behavior of the transmission in the ultra-high-angle range, between \ang{80} to \ang{90}, shows the difficulty of designing high-NA edge-detection devices in transmission mode: \textit{any} multilayer dielectric film will tend towards perfect reflection at glancing incidence (\ang{90}), in which case the design of a quadratic transmission profile that increases from 0 to a maximum at, say, \ang{80}, is highly unnatural and hard to achieve. Conversely, designing reflection-mode designs is significantly simpler. Yet transmission is the ideal operational mode for high-speed edge detection, and the designs presented in \figref{figure2} achieve this with high fidelity at high NA (0.98). The intensity of the output field is shown in \figref{figure2}(d), where one can see that the edges, oriented in all directions, are clearly resolvable.

\begin{figure*}[!hbt]
    \includegraphics[width=\textwidth]{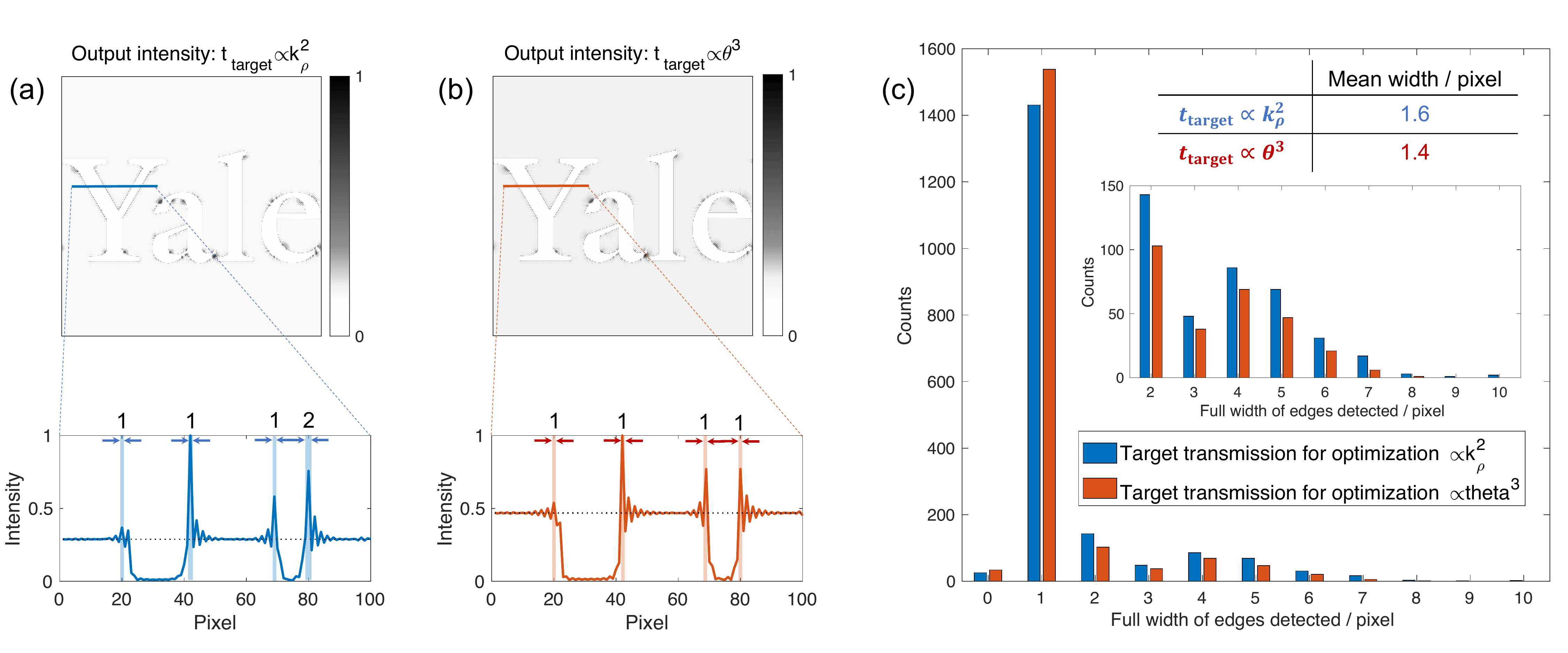}
    \centering
    \caption{Width distributions of the detected edges for the two target transmission coefficients depicted in \figref{figure2}. (a,b) Output images (top) and distributions of edges along an example line (bottom) for (a) $t_{\rm target}\propto{k_{\rho}}^2$ and (b) $t_{\rm target}\propto{\theta}^3$, respectively. The edge widths are defined as the widths over which the intensity surpass the noise level. (c) Histogram of widths in each approach, showing that the cubic-in-angle target coefficient yields slightly better results, i.e., narrower edges, than the quadratic-in-wavevector target coefficient. Non-quadratic target functions can be superior when accounting for the scattering properties of real physical implementations.}
    \label{fig:figure3}
\end{figure*}

As mentioned in the introduction, edge detection requires good filtering of the spatial frequencies of the incoming wave, but such filtering does \emph{not} necessarily require the $\sim k_{\rho}^2$ dependence of the Laplacian operator. The discrepancy between the optimal-design transmission coefficient and that of the target in \figref{figure2}(c) suggests that it may be impossible for multilayer structures to achieve perfect quadratic scaling of their transmission coefficients alongside minimal phase variations, which would imply that the optimal design of \figref{figure2}(b) may have paid some penalty in attempting to minimize the difference with a quadratic target, instead of simply aiming for good filtering properties.

In \figref{figure2}(e), we show an alternative design that emerged for a target transmission coefficient given by $t_{\rm target}(\theta) = \alpha \theta^3$ (cf. SM). We chose this function to more closely match the transmission curves of real multilayer designs, and \figref{figure2}(f) shows the much closer match between the transmission coefficient of the new optimal design and that of the new target. In \figref{figure2}(g) the output-field intensity again shows very good edge resolution, although it is difficult by eye to detect which of the designs of \figref{figure2}(b,e) is better.

\Figref{figure3} quantitatively compares the edge image quality of the two designs of \figref{figure2}(b,e). We measure the width of every edge that is present in the output, and compare between the two designs (also using the ground truth that is known from the input image). \Figref{figure3}(a,b) demonstrates prototypical results, where the design using $k_{\rho}^2$ target transmission exhibits thicker edges, as measured by the number of pixels above the background intensity (dashed lines), than its counterpart designed with $t_{\rm target} \sim \theta^3$. The histogram of \figref{figure3}(c) shows the relative numbers of edges with a given width (in terms of number of pixels) for the two designs, with the $\theta^3$ design offering slightly better performance and a smaller average width per pixel. (The number of edges missed entirely is nonzero but very small for both designs.) This demonstrates that, although the Laplacian operator may be a good starting point for edge-detection design, it is neither required nor necessarily globally optimal, which is likely true for alternative edge-detection approaches (metasurface, plasmonic, etc.) as well.

\section{Discussions}
\begin{figure}[!hbt]
    \includegraphics[width=\linewidth]{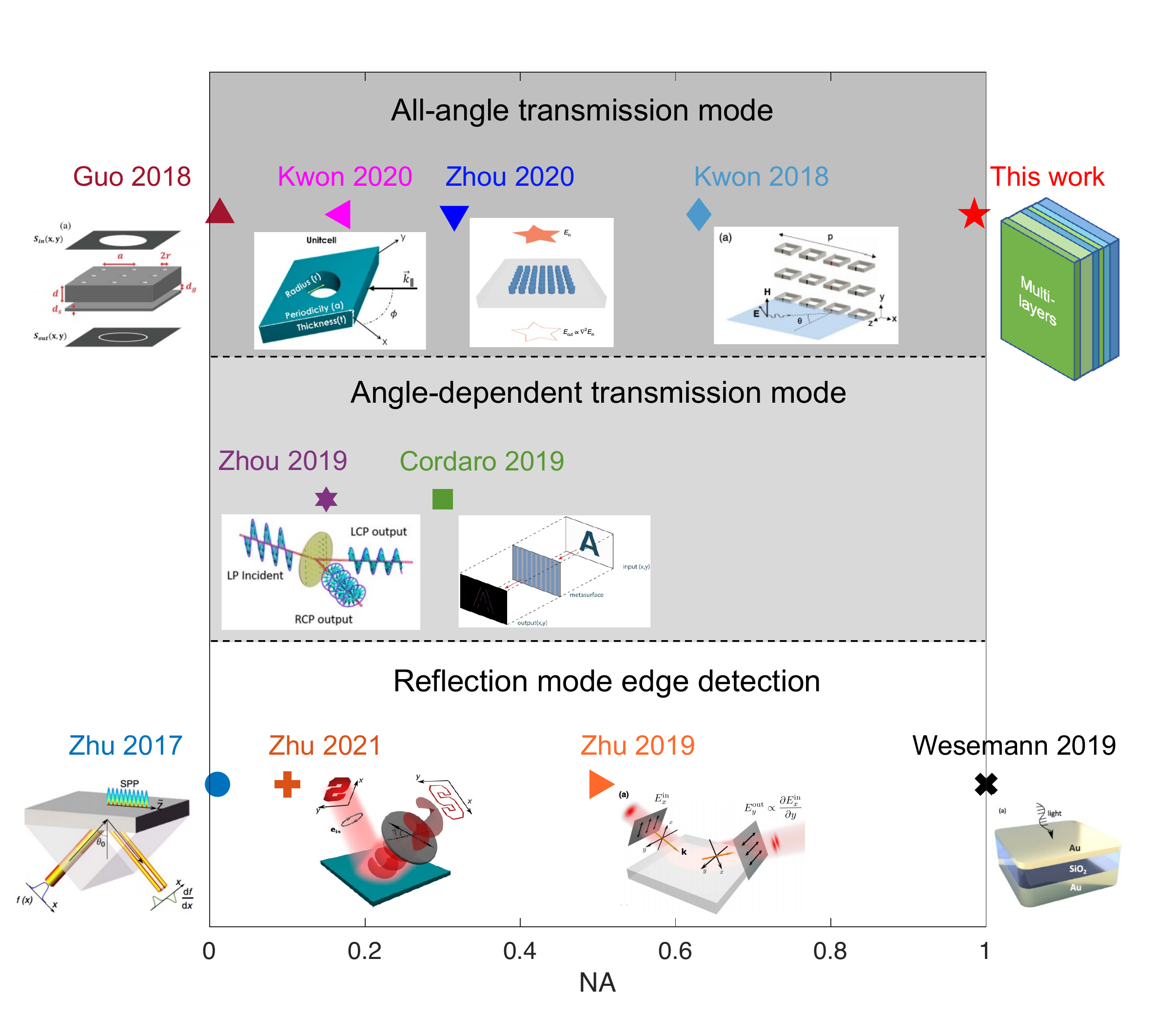}
    \centering
    \caption{Comparison of the multilayer designs to alternative, recently proposed designs~\cite{zhu2017plasmonic,guo2018photonic,kwon2018nonlocal,zhu2019generalized,zhou2019optical,wesemann2019selective,cordaro2019high,zhou2020flat,kwon2020dual,zhu2021topological}. The top dark-grey region indicates structures that effectively detect edges in two-dimensional images in transmission mode, while structures in the middle light-grey region achieve partial-2D detection or detect edges along one-dimensional lines. Structures in the bottom white region work in reflection mode. Multilayer-designs show uniquely strong capabilities along the three key features of two-dimensional image formation, high numerical aperture (0.98 in this work), and feasibility of fabrication.}
    \label{fig:figure4}
\end{figure}

In \figref{figure4} we compare the multilayer designs of \figref{figure2}(b,e) to other recent state-of-the-art designs~\cite{zhu2017plasmonic,guo2018photonic,kwon2018nonlocal,zhu2019generalized,zhou2019optical,wesemann2019selective,cordaro2019high,zhou2020flat,kwon2020dual,zhu2021topological}. The designs in the bottom region (white) of the figure operate in reflection mode, which is easier to design but not ideal to implement. The design denoted by the black cross is multilayer-film-based, but it has large phase variations in the transmission coefficient as a function of angle ($> \ang{180}$), which significantly blurs the edge resolution (cf. SM). The designs in the middle region (light grey) work in transmission mode but only for limited operation across all possible azimuthal angles. (We also note that the metasurface design of \citeasnoun{zhou2019optical}, denoted by a purple hexagon, requires lenses and thus does not offer space savings.) The designs in the top region (dark grey) all work in transmission mode and have rotational symmetry, under continuous or discrete rotations, that make their scattering response independent (or nearly so) of azimuthal angle. Of these designs, three~\cite{guo2018photonic,kwon2020dual,zhou2020flat} operate only for low numerical aperature (which corresponds to operation over only a narrow range of wavevectors), while the fourth~\cite{kwon2018nonlocal} is difficult or impossible to fabricate. The optimized multilayer films of this work show a clear advantage along these three dimensions.

Finally, we analyze the effects of fabrication errors on the designed structures. We simulate random errors in the layer thicknesses by sampling from the normalized Gaussian distribution $\mathcal{N}(0,\sigma)$, where 0 represents the mean shift from the desired thicknesses and $\sigma$ represents the standard deviation. We find that for the design with $t_{\rm target}\propto k_{\rho}^2$, the objective function in \eqref{opt_prob} for $p$-polarization increases on average by at most $25\%$ when $\sigma\leq\SI{3}{nm}$, while for $s$-polarization the transmission hardly changes from 0 at all angles. For the design with $t_{\rm target}\propto {\theta}^3$, the equivalent deviation is $\sigma\leq\SI{1.8}{nm}$ for $25\%$ error. The performance error varies smoothly with the fabrication error, and such tolerances are well beyond the angstrom-level thickness errors in state-of-the-art LPCVD fabrication~\cite{joubert1987effect,voutsas1992structure,yang2000new}. Alternative, easier-to-synthesize materials can be designed using the same techniques described above and achieve similar performance.

\section{Outlook}
Looking forward, computational optimization of multilayer structures may enable a wide range of analog linear operators, beyond just edge detection. Instead of defining specific target transmission coefficient profiles, as in \eqref{opt_prob}, one could utilize a data-driven approach that matches the desired features in a given scattered field with those of known image/field pairs. We implemented such an approach specifically for edge detection, but the performance of the optimal designs was nearly equivalent to the best designs already shown here. Another possible direction to explore is towards significantly thicker multilayer designs, which may enable efficient multi-frequency performance, though in such a case local-optimization techniques might falter and require global-optimization techniques instead.

\emph{Acknowledgments}---The authors thank Francesco Monticone for useful discussions. This work was supported by the Air Force Office of Scientific Research under award number FA9550-17-1-009.
\bibliography{edge_bib}

%apsrev4-2.bst 2019-01-14 (MD) hand-edited version of apsrev4-1.bst
%Control: key (0)
%Control: author (8) initials jnrlst
%Control: editor formatted (1) identically to author
%Control: production of article title (0) allowed
%Control: page (0) single
%Control: year (1) truncated
%Control: production of eprint (0) enabled
\begin{thebibliography}{30}%
\makeatletter
\providecommand \@ifxundefined [1]{%
 \@ifx{#1\undefined}
}%
\providecommand \@ifnum [1]{%
 \ifnum #1\expandafter \@firstoftwo
 \else \expandafter \@secondoftwo
 \fi
}%
\providecommand \@ifx [1]{%
 \ifx #1\expandafter \@firstoftwo
 \else \expandafter \@secondoftwo
 \fi
}%
\providecommand \natexlab [1]{#1}%
\providecommand \enquote  [1]{``#1''}%
\providecommand \bibnamefont  [1]{#1}%
\providecommand \bibfnamefont [1]{#1}%
\providecommand \citenamefont [1]{#1}%
\providecommand \href@noop [0]{\@secondoftwo}%
\providecommand \href [0]{\begingroup \@sanitize@url \@href}%
\providecommand \@href[1]{\@@startlink{#1}\@@href}%
\providecommand \@@href[1]{\endgroup#1\@@endlink}%
\providecommand \@sanitize@url [0]{\catcode `\\12\catcode `\$12\catcode
  `\&12\catcode `\#12\catcode `\^12\catcode `\_12\catcode `\%12\relax}%
\providecommand \@@startlink[1]{}%
\providecommand \@@endlink[0]{}%
\providecommand \url  [0]{\begingroup\@sanitize@url \@url }%
\providecommand \@url [1]{\endgroup\@href {#1}{\urlprefix }}%
\providecommand \urlprefix  [0]{URL }%
\providecommand \Eprint [0]{\href }%
\providecommand \doibase [0]{https://doi.org/}%
\providecommand \selectlanguage [0]{\@gobble}%
\providecommand \bibinfo  [0]{\@secondoftwo}%
\providecommand \bibfield  [0]{\@secondoftwo}%
\providecommand \translation [1]{[#1]}%
\providecommand \BibitemOpen [0]{}%
\providecommand \bibitemStop [0]{}%
\providecommand \bibitemNoStop [0]{.\EOS\space}%
\providecommand \EOS [0]{\spacefactor3000\relax}%
\providecommand \BibitemShut  [1]{\csname bibitem#1\endcsname}%
\let\auto@bib@innerbib\@empty
%</preamble>
\bibitem [{\citenamefont {Solli}\ and\ \citenamefont
  {Jalali}(2015)}]{solli2015analog}%
  \BibitemOpen
  \bibfield  {author} {\bibinfo {author} {\bibfnamefont {D.~R.}\ \bibnamefont
  {Solli}}\ and\ \bibinfo {author} {\bibfnamefont {B.}~\bibnamefont {Jalali}},\
  }\bibfield  {title} {\bibinfo {title} {Analog optical computing},\ }\href
  {https://doi.org/10.1038/nphoton.2015.208} {\bibfield  {journal} {\bibinfo
  {journal} {Nature Photonics}\ }\textbf {\bibinfo {volume} {9}},\ \bibinfo
  {pages} {704} (\bibinfo {year} {2015})}\BibitemShut {NoStop}%
\bibitem [{\citenamefont {Zhu}\ \emph {et~al.}(2017)\citenamefont {Zhu},
  \citenamefont {Zhou}, \citenamefont {Lou}, \citenamefont {Ye}, \citenamefont
  {Qiu}, \citenamefont {Ruan},\ and\ \citenamefont {Fan}}]{zhu2017plasmonic}%
  \BibitemOpen
  \bibfield  {author} {\bibinfo {author} {\bibfnamefont {T.}~\bibnamefont
  {Zhu}}, \bibinfo {author} {\bibfnamefont {Y.}~\bibnamefont {Zhou}}, \bibinfo
  {author} {\bibfnamefont {Y.}~\bibnamefont {Lou}}, \bibinfo {author}
  {\bibfnamefont {H.}~\bibnamefont {Ye}}, \bibinfo {author} {\bibfnamefont
  {M.}~\bibnamefont {Qiu}}, \bibinfo {author} {\bibfnamefont {Z.}~\bibnamefont
  {Ruan}},\ and\ \bibinfo {author} {\bibfnamefont {S.}~\bibnamefont {Fan}},\
  }\bibfield  {title} {\bibinfo {title} {Plasmonic computing of spatial
  differentiation},\ }\href {https://doi.org/10.1038/ncomms15391} {\bibfield
  {journal} {\bibinfo  {journal} {Nature communications}\ }\textbf {\bibinfo
  {volume} {8}},\ \bibinfo {pages} {1} (\bibinfo {year} {2017})}\BibitemShut
  {NoStop}%
\bibitem [{\citenamefont {Guo}\ \emph {et~al.}(2018)\citenamefont {Guo},
  \citenamefont {Xiao}, \citenamefont {Minkov}, \citenamefont {Shi},\ and\
  \citenamefont {Fan}}]{guo2018photonic}%
  \BibitemOpen
  \bibfield  {author} {\bibinfo {author} {\bibfnamefont {C.}~\bibnamefont
  {Guo}}, \bibinfo {author} {\bibfnamefont {M.}~\bibnamefont {Xiao}}, \bibinfo
  {author} {\bibfnamefont {M.}~\bibnamefont {Minkov}}, \bibinfo {author}
  {\bibfnamefont {Y.}~\bibnamefont {Shi}},\ and\ \bibinfo {author}
  {\bibfnamefont {S.}~\bibnamefont {Fan}},\ }\bibfield  {title} {\bibinfo
  {title} {Photonic crystal slab laplace operator for image differentiation},\
  }\href {https://doi.org/10.1364/OPTICA.5.000251} {\bibfield  {journal}
  {\bibinfo  {journal} {Optica}\ }\textbf {\bibinfo {volume} {5}},\ \bibinfo
  {pages} {251} (\bibinfo {year} {2018})}\BibitemShut {NoStop}%
\bibitem [{\citenamefont {Kwon}\ \emph {et~al.}(2018)\citenamefont {Kwon},
  \citenamefont {Sounas}, \citenamefont {Cordaro}, \citenamefont {Polman},\
  and\ \citenamefont {Al{\`u}}}]{kwon2018nonlocal}%
  \BibitemOpen
  \bibfield  {author} {\bibinfo {author} {\bibfnamefont {H.}~\bibnamefont
  {Kwon}}, \bibinfo {author} {\bibfnamefont {D.}~\bibnamefont {Sounas}},
  \bibinfo {author} {\bibfnamefont {A.}~\bibnamefont {Cordaro}}, \bibinfo
  {author} {\bibfnamefont {A.}~\bibnamefont {Polman}},\ and\ \bibinfo {author}
  {\bibfnamefont {A.}~\bibnamefont {Al{\`u}}},\ }\bibfield  {title} {\bibinfo
  {title} {Nonlocal metasurfaces for optical signal processing},\ }\href
  {https://doi.org/10.1103/PhysRevLett.121.173004} {\bibfield  {journal}
  {\bibinfo  {journal} {Physical review letters}\ }\textbf {\bibinfo {volume}
  {121}},\ \bibinfo {pages} {173004} (\bibinfo {year} {2018})}\BibitemShut
  {NoStop}%
\bibitem [{\citenamefont {Zhu}\ \emph {et~al.}(2019)\citenamefont {Zhu},
  \citenamefont {Lou}, \citenamefont {Zhou}, \citenamefont {Zhang},
  \citenamefont {Huang}, \citenamefont {Li}, \citenamefont {Luo}, \citenamefont
  {Wen}, \citenamefont {Zhu}, \citenamefont {Gong} \emph
  {et~al.}}]{zhu2019generalized}%
  \BibitemOpen
  \bibfield  {author} {\bibinfo {author} {\bibfnamefont {T.}~\bibnamefont
  {Zhu}}, \bibinfo {author} {\bibfnamefont {Y.}~\bibnamefont {Lou}}, \bibinfo
  {author} {\bibfnamefont {Y.}~\bibnamefont {Zhou}}, \bibinfo {author}
  {\bibfnamefont {J.}~\bibnamefont {Zhang}}, \bibinfo {author} {\bibfnamefont
  {J.}~\bibnamefont {Huang}}, \bibinfo {author} {\bibfnamefont
  {Y.}~\bibnamefont {Li}}, \bibinfo {author} {\bibfnamefont {H.}~\bibnamefont
  {Luo}}, \bibinfo {author} {\bibfnamefont {S.}~\bibnamefont {Wen}}, \bibinfo
  {author} {\bibfnamefont {S.}~\bibnamefont {Zhu}}, \bibinfo {author}
  {\bibfnamefont {Q.}~\bibnamefont {Gong}}, \emph {et~al.},\ }\bibfield
  {title} {\bibinfo {title} {Generalized spatial differentiation from the spin
  hall effect of light and its application in image processing of edge
  detection},\ }\href {https://doi.org/10.1103/PhysRevApplied.11.034043}
  {\bibfield  {journal} {\bibinfo  {journal} {Physical Review Applied}\
  }\textbf {\bibinfo {volume} {11}},\ \bibinfo {pages} {034043} (\bibinfo
  {year} {2019})}\BibitemShut {NoStop}%
\bibitem [{\citenamefont {Zhou}\ \emph {et~al.}(2019)\citenamefont {Zhou},
  \citenamefont {Qian}, \citenamefont {Chen}, \citenamefont {Zhao},
  \citenamefont {Li}, \citenamefont {Wu}, \citenamefont {Luo}, \citenamefont
  {Wen},\ and\ \citenamefont {Liu}}]{zhou2019optical}%
  \BibitemOpen
  \bibfield  {author} {\bibinfo {author} {\bibfnamefont {J.}~\bibnamefont
  {Zhou}}, \bibinfo {author} {\bibfnamefont {H.}~\bibnamefont {Qian}}, \bibinfo
  {author} {\bibfnamefont {C.-F.}\ \bibnamefont {Chen}}, \bibinfo {author}
  {\bibfnamefont {J.}~\bibnamefont {Zhao}}, \bibinfo {author} {\bibfnamefont
  {G.}~\bibnamefont {Li}}, \bibinfo {author} {\bibfnamefont {Q.}~\bibnamefont
  {Wu}}, \bibinfo {author} {\bibfnamefont {H.}~\bibnamefont {Luo}}, \bibinfo
  {author} {\bibfnamefont {S.}~\bibnamefont {Wen}},\ and\ \bibinfo {author}
  {\bibfnamefont {Z.}~\bibnamefont {Liu}},\ }\bibfield  {title} {\bibinfo
  {title} {Optical edge detection based on high-efficiency dielectric
  metasurface},\ }\href {https://doi.org/10.1073/pnas.1820636116} {\bibfield
  {journal} {\bibinfo  {journal} {Proceedings of the National Academy of
  Sciences}\ }\textbf {\bibinfo {volume} {116}},\ \bibinfo {pages} {11137}
  (\bibinfo {year} {2019})}\BibitemShut {NoStop}%
\bibitem [{\citenamefont {Wesemann}\ \emph {et~al.}(2019)\citenamefont
  {Wesemann}, \citenamefont {Panchenko}, \citenamefont {Singh}, \citenamefont
  {Della~Gaspera}, \citenamefont {G{\'o}mez}, \citenamefont {Davis},\ and\
  \citenamefont {Roberts}}]{wesemann2019selective}%
  \BibitemOpen
  \bibfield  {author} {\bibinfo {author} {\bibfnamefont {L.}~\bibnamefont
  {Wesemann}}, \bibinfo {author} {\bibfnamefont {E.}~\bibnamefont {Panchenko}},
  \bibinfo {author} {\bibfnamefont {K.}~\bibnamefont {Singh}}, \bibinfo
  {author} {\bibfnamefont {E.}~\bibnamefont {Della~Gaspera}}, \bibinfo {author}
  {\bibfnamefont {D.~E.}\ \bibnamefont {G{\'o}mez}}, \bibinfo {author}
  {\bibfnamefont {T.~J.}\ \bibnamefont {Davis}},\ and\ \bibinfo {author}
  {\bibfnamefont {A.}~\bibnamefont {Roberts}},\ }\bibfield  {title} {\bibinfo
  {title} {Selective near-perfect absorbing mirror as a spatial frequency
  filter for optical image processing},\ }\href@noop {} {\bibfield  {journal}
  {\bibinfo  {journal} {APL Photonics}\ }\textbf {\bibinfo {volume} {4}},\
  \bibinfo {pages} {100801} (\bibinfo {year} {2019})}\BibitemShut {NoStop}%
\bibitem [{\citenamefont {Cordaro}\ \emph {et~al.}(2019)\citenamefont
  {Cordaro}, \citenamefont {Kwon}, \citenamefont {Sounas}, \citenamefont
  {Koenderink}, \citenamefont {Al{\`u}},\ and\ \citenamefont
  {Polman}}]{cordaro2019high}%
  \BibitemOpen
  \bibfield  {author} {\bibinfo {author} {\bibfnamefont {A.}~\bibnamefont
  {Cordaro}}, \bibinfo {author} {\bibfnamefont {H.}~\bibnamefont {Kwon}},
  \bibinfo {author} {\bibfnamefont {D.}~\bibnamefont {Sounas}}, \bibinfo
  {author} {\bibfnamefont {A.~F.}\ \bibnamefont {Koenderink}}, \bibinfo
  {author} {\bibfnamefont {A.}~\bibnamefont {Al{\`u}}},\ and\ \bibinfo {author}
  {\bibfnamefont {A.}~\bibnamefont {Polman}},\ }\bibfield  {title} {\bibinfo
  {title} {High-index dielectric metasurfaces performing mathematical
  operations},\ }\href {https://doi.org/10.1021/acs.nanolett.9b02477}
  {\bibfield  {journal} {\bibinfo  {journal} {Nano letters}\ }\textbf {\bibinfo
  {volume} {19}},\ \bibinfo {pages} {8418} (\bibinfo {year}
  {2019})}\BibitemShut {NoStop}%
\bibitem [{\citenamefont {Zhou}\ \emph {et~al.}(2020)\citenamefont {Zhou},
  \citenamefont {Zheng}, \citenamefont {Kravchenko},\ and\ \citenamefont
  {Valentine}}]{zhou2020flat}%
  \BibitemOpen
  \bibfield  {author} {\bibinfo {author} {\bibfnamefont {Y.}~\bibnamefont
  {Zhou}}, \bibinfo {author} {\bibfnamefont {H.}~\bibnamefont {Zheng}},
  \bibinfo {author} {\bibfnamefont {I.~I.}\ \bibnamefont {Kravchenko}},\ and\
  \bibinfo {author} {\bibfnamefont {J.}~\bibnamefont {Valentine}},\ }\bibfield
  {title} {\bibinfo {title} {Flat optics for image differentiation},\ }\href
  {https://doi.org/10.1038/s41566-020-0591-3} {\bibfield  {journal} {\bibinfo
  {journal} {Nature Photonics}\ }\textbf {\bibinfo {volume} {14}},\ \bibinfo
  {pages} {316} (\bibinfo {year} {2020})}\BibitemShut {NoStop}%
\bibitem [{\citenamefont {Kwon}\ \emph {et~al.}(2020)\citenamefont {Kwon},
  \citenamefont {Cordaro}, \citenamefont {Sounas}, \citenamefont {Polman},\
  and\ \citenamefont {Alu}}]{kwon2020dual}%
  \BibitemOpen
  \bibfield  {author} {\bibinfo {author} {\bibfnamefont {H.}~\bibnamefont
  {Kwon}}, \bibinfo {author} {\bibfnamefont {A.}~\bibnamefont {Cordaro}},
  \bibinfo {author} {\bibfnamefont {D.}~\bibnamefont {Sounas}}, \bibinfo
  {author} {\bibfnamefont {A.}~\bibnamefont {Polman}},\ and\ \bibinfo {author}
  {\bibfnamefont {A.}~\bibnamefont {Alu}},\ }\bibfield  {title} {\bibinfo
  {title} {Dual-polarization analog 2d image processing with nonlocal
  metasurfaces},\ }\href {https://doi.org/10.1021/acsphotonics.0c00473}
  {\bibfield  {journal} {\bibinfo  {journal} {ACS Photonics}\ }\textbf
  {\bibinfo {volume} {7}},\ \bibinfo {pages} {1799} (\bibinfo {year}
  {2020})}\BibitemShut {NoStop}%
\bibitem [{\citenamefont {Zhu}\ \emph {et~al.}(2021)\citenamefont {Zhu},
  \citenamefont {Guo}, \citenamefont {Huang}, \citenamefont {Wang},
  \citenamefont {Orenstein}, \citenamefont {Ruan},\ and\ \citenamefont
  {Fan}}]{zhu2021topological}%
  \BibitemOpen
  \bibfield  {author} {\bibinfo {author} {\bibfnamefont {T.}~\bibnamefont
  {Zhu}}, \bibinfo {author} {\bibfnamefont {C.}~\bibnamefont {Guo}}, \bibinfo
  {author} {\bibfnamefont {J.}~\bibnamefont {Huang}}, \bibinfo {author}
  {\bibfnamefont {H.}~\bibnamefont {Wang}}, \bibinfo {author} {\bibfnamefont
  {M.}~\bibnamefont {Orenstein}}, \bibinfo {author} {\bibfnamefont
  {Z.}~\bibnamefont {Ruan}},\ and\ \bibinfo {author} {\bibfnamefont
  {S.}~\bibnamefont {Fan}},\ }\bibfield  {title} {\bibinfo {title} {Topological
  optical differentiator},\ }\href@noop {} {\bibfield  {journal} {\bibinfo
  {journal} {Nature communications}\ }\textbf {\bibinfo {volume} {12}},\
  \bibinfo {pages} {1} (\bibinfo {year} {2021})}\BibitemShut {NoStop}%
\bibitem [{\citenamefont {Yeh}\ \emph {et~al.}(1988)\citenamefont {Yeh} \emph
  {et~al.}}]{yeh1988optical}%
  \BibitemOpen
  \bibfield  {author} {\bibinfo {author} {\bibfnamefont {P.}~\bibnamefont
  {Yeh}} \emph {et~al.},\ }\href@noop {} {\emph {\bibinfo {title} {Optical
  waves in layered media}}},\ Vol.~\bibinfo {volume} {95}\ (\bibinfo
  {publisher} {Wiley New York},\ \bibinfo {year} {1988})\BibitemShut {NoStop}%
\bibitem [{\citenamefont {Faist}\ \emph {et~al.}(1994)\citenamefont {Faist},
  \citenamefont {Capasso}, \citenamefont {Sivco}, \citenamefont {Sirtori},
  \citenamefont {Hutchinson},\ and\ \citenamefont {Cho}}]{faist1994quantum}%
  \BibitemOpen
  \bibfield  {author} {\bibinfo {author} {\bibfnamefont {J.}~\bibnamefont
  {Faist}}, \bibinfo {author} {\bibfnamefont {F.}~\bibnamefont {Capasso}},
  \bibinfo {author} {\bibfnamefont {D.~L.}\ \bibnamefont {Sivco}}, \bibinfo
  {author} {\bibfnamefont {C.}~\bibnamefont {Sirtori}}, \bibinfo {author}
  {\bibfnamefont {A.~L.}\ \bibnamefont {Hutchinson}},\ and\ \bibinfo {author}
  {\bibfnamefont {A.~Y.}\ \bibnamefont {Cho}},\ }\bibfield  {title} {\bibinfo
  {title} {Quantum cascade laser},\ }\href
  {https://doi.org/10.1126/science.264.5158.553} {\bibfield  {journal}
  {\bibinfo  {journal} {Science}\ }\textbf {\bibinfo {volume} {264}},\ \bibinfo
  {pages} {553} (\bibinfo {year} {1994})}\BibitemShut {NoStop}%
\bibitem [{\citenamefont {Silva}\ \emph {et~al.}(2014)\citenamefont {Silva},
  \citenamefont {Monticone}, \citenamefont {Castaldi}, \citenamefont {Galdi},
  \citenamefont {Al{\`u}},\ and\ \citenamefont
  {Engheta}}]{silva2014performing}%
  \BibitemOpen
  \bibfield  {author} {\bibinfo {author} {\bibfnamefont {A.}~\bibnamefont
  {Silva}}, \bibinfo {author} {\bibfnamefont {F.}~\bibnamefont {Monticone}},
  \bibinfo {author} {\bibfnamefont {G.}~\bibnamefont {Castaldi}}, \bibinfo
  {author} {\bibfnamefont {V.}~\bibnamefont {Galdi}}, \bibinfo {author}
  {\bibfnamefont {A.}~\bibnamefont {Al{\`u}}},\ and\ \bibinfo {author}
  {\bibfnamefont {N.}~\bibnamefont {Engheta}},\ }\bibfield  {title} {\bibinfo
  {title} {Performing mathematical operations with metamaterials},\ }\href
  {https://doi.org/10.1126/science.1242818} {\bibfield  {journal} {\bibinfo
  {journal} {Science}\ }\textbf {\bibinfo {volume} {343}},\ \bibinfo {pages}
  {160} (\bibinfo {year} {2014})}\BibitemShut {NoStop}%
\bibitem [{\citenamefont {Lissberger}(1970)}]{lissberger1970optical}%
  \BibitemOpen
  \bibfield  {author} {\bibinfo {author} {\bibfnamefont {P.~H.}\ \bibnamefont
  {Lissberger}},\ }\bibfield  {title} {\bibinfo {title} {Optical applications
  of dielectric thin films},\ }\href
  {https://doi.org/10.1088/0034-4885/33/1/305} {\bibfield  {journal} {\bibinfo
  {journal} {Reports on Progress in physics}\ }\textbf {\bibinfo {volume}
  {33}},\ \bibinfo {pages} {197} (\bibinfo {year} {1970})}\BibitemShut
  {NoStop}%
\bibitem [{\citenamefont {Goodman}(2005)}]{goodman2005introduction}%
  \BibitemOpen
  \bibfield  {author} {\bibinfo {author} {\bibfnamefont {J.~W.}\ \bibnamefont
  {Goodman}},\ }\href@noop {} {\emph {\bibinfo {title} {Introduction to Fourier
  optics}}}\ (\bibinfo  {publisher} {Roberts and Company Publishers},\ \bibinfo
  {year} {2005})\BibitemShut {NoStop}%
\bibitem [{\citenamefont {Jaeger}(2002)}]{jaeger2002introduction}%
  \BibitemOpen
  \bibfield  {author} {\bibinfo {author} {\bibfnamefont {R.~C.}\ \bibnamefont
  {Jaeger}},\ }\href@noop {} {\emph {\bibinfo {title} {Introduction to
  microelectronic fabrication}}},\ Vol.~\bibinfo {volume} {2}\ (\bibinfo
  {publisher} {Prentice Hall Upper Saddle River, NJ},\ \bibinfo {year}
  {2002})\BibitemShut {NoStop}%
\bibitem [{\citenamefont {Seshan}(2001)}]{seshan2001handbook}%
  \BibitemOpen
  \bibfield  {author} {\bibinfo {author} {\bibfnamefont {K.}~\bibnamefont
  {Seshan}},\ }\href@noop {} {\emph {\bibinfo {title} {Handbook of thin film
  deposition processes and techniques}}}\ (\bibinfo  {publisher} {William
  Andrew},\ \bibinfo {year} {2001})\BibitemShut {NoStop}%
\bibitem [{\citenamefont {vd~Laan}\ and\ \citenamefont
  {Frankena}(1978)}]{vd1978fast}%
  \BibitemOpen
  \bibfield  {author} {\bibinfo {author} {\bibfnamefont {C.}~\bibnamefont
  {vd~Laan}}\ and\ \bibinfo {author} {\bibfnamefont {H.}~\bibnamefont
  {Frankena}},\ }\bibfield  {title} {\bibinfo {title} {Fast computation method
  for derivatives of multilayer stack reflectance},\ }\href
  {https://doi.org/10.1364/AO.17.000538} {\bibfield  {journal} {\bibinfo
  {journal} {Applied Optics}\ }\textbf {\bibinfo {volume} {17}},\ \bibinfo
  {pages} {538} (\bibinfo {year} {1978})}\BibitemShut {NoStop}%
\bibitem [{\citenamefont {Peng}\ and\ \citenamefont
  {Marcel}(1985)}]{peng1985derivatives}%
  \BibitemOpen
  \bibfield  {author} {\bibinfo {author} {\bibfnamefont {K.-O.}\ \bibnamefont
  {Peng}}\ and\ \bibinfo {author} {\bibfnamefont {R.}~\bibnamefont {Marcel}},\
  }\bibfield  {title} {\bibinfo {title} {Derivatives of transmittance and
  reflectance for an absorbing multilayer stack},\ }\href
  {https://doi.org/10.1364/AO.24.000501} {\bibfield  {journal} {\bibinfo
  {journal} {Applied optics}\ }\textbf {\bibinfo {volume} {24}},\ \bibinfo
  {pages} {501} (\bibinfo {year} {1985})}\BibitemShut {NoStop}%
\bibitem [{\citenamefont {Tikhonravov}(1982)}]{tikhonravov1982synthesis}%
  \BibitemOpen
  \bibfield  {author} {\bibinfo {author} {\bibfnamefont {A.}~\bibnamefont
  {Tikhonravov}},\ }\bibfield  {title} {\bibinfo {title} {Synthesis of optical
  coatings using optimality conditions},\ }\href@noop {} {\bibfield  {journal}
  {\bibinfo  {journal} {Vestnik MGU, physics and astronomy series}\ }\textbf
  {\bibinfo {volume} {23}},\ \bibinfo {pages} {91} (\bibinfo {year}
  {1982})}\BibitemShut {NoStop}%
\bibitem [{\citenamefont {Tikhonravov}\ \emph {et~al.}(1996)\citenamefont
  {Tikhonravov}, \citenamefont {Trubetskov},\ and\ \citenamefont
  {DeBell}}]{tikhonravov1996application}%
  \BibitemOpen
  \bibfield  {author} {\bibinfo {author} {\bibfnamefont {A.~V.}\ \bibnamefont
  {Tikhonravov}}, \bibinfo {author} {\bibfnamefont {M.~K.}\ \bibnamefont
  {Trubetskov}},\ and\ \bibinfo {author} {\bibfnamefont {G.~W.}\ \bibnamefont
  {DeBell}},\ }\bibfield  {title} {\bibinfo {title} {Application of the needle
  optimization technique to the design of optical coatings},\ }\href
  {https://doi.org/10.1364/AO.35.005493} {\bibfield  {journal} {\bibinfo
  {journal} {Applied optics}\ }\textbf {\bibinfo {volume} {35}},\ \bibinfo
  {pages} {5493} (\bibinfo {year} {1996})}\BibitemShut {NoStop}%
\bibitem [{\citenamefont {Shim}\ \emph {et~al.}(2020)\citenamefont {Shim},
  \citenamefont {Kuang},\ and\ \citenamefont {Miller}}]{shim2020optical}%
  \BibitemOpen
  \bibfield  {author} {\bibinfo {author} {\bibfnamefont {H.}~\bibnamefont
  {Shim}}, \bibinfo {author} {\bibfnamefont {Z.}~\bibnamefont {Kuang}},\ and\
  \bibinfo {author} {\bibfnamefont {O.~D.}\ \bibnamefont {Miller}},\ }\bibfield
   {title} {\bibinfo {title} {Optical materials for maximal nanophotonic
  response},\ }\href@noop {} {\bibfield  {journal} {\bibinfo  {journal}
  {Optical Materials Express}\ }\textbf {\bibinfo {volume} {10}},\ \bibinfo
  {pages} {1561} (\bibinfo {year} {2020})}\BibitemShut {NoStop}%
\bibitem [{\citenamefont {Byrd}\ \emph {et~al.}(2000)\citenamefont {Byrd},
  \citenamefont {Gilbert},\ and\ \citenamefont {Nocedal}}]{byrd2000trust}%
  \BibitemOpen
  \bibfield  {author} {\bibinfo {author} {\bibfnamefont {R.~H.}\ \bibnamefont
  {Byrd}}, \bibinfo {author} {\bibfnamefont {J.~C.}\ \bibnamefont {Gilbert}},\
  and\ \bibinfo {author} {\bibfnamefont {J.}~\bibnamefont {Nocedal}},\
  }\bibfield  {title} {\bibinfo {title} {A trust region method based on
  interior point techniques for nonlinear programming},\ }\href
  {https://doi.org/10.1007/PL00011391} {\bibfield  {journal} {\bibinfo
  {journal} {Mathematical programming}\ }\textbf {\bibinfo {volume} {89}},\
  \bibinfo {pages} {149} (\bibinfo {year} {2000})}\BibitemShut {NoStop}%
\bibitem [{\citenamefont {Byrd}\ \emph {et~al.}(1999)\citenamefont {Byrd},
  \citenamefont {Hribar},\ and\ \citenamefont {Nocedal}}]{byrd1999interior}%
  \BibitemOpen
  \bibfield  {author} {\bibinfo {author} {\bibfnamefont {R.~H.}\ \bibnamefont
  {Byrd}}, \bibinfo {author} {\bibfnamefont {M.~E.}\ \bibnamefont {Hribar}},\
  and\ \bibinfo {author} {\bibfnamefont {J.}~\bibnamefont {Nocedal}},\
  }\bibfield  {title} {\bibinfo {title} {An interior point algorithm for
  large-scale nonlinear programming},\ }\href
  {https://doi.org/10.1137/S1052623497325107} {\bibfield  {journal} {\bibinfo
  {journal} {SIAM Journal on Optimization}\ }\textbf {\bibinfo {volume} {9}},\
  \bibinfo {pages} {877} (\bibinfo {year} {1999})}\BibitemShut {NoStop}%
\bibitem [{\citenamefont {Waltz}\ \emph {et~al.}(2006)\citenamefont {Waltz},
  \citenamefont {Morales}, \citenamefont {Nocedal},\ and\ \citenamefont
  {Orban}}]{waltz2006interior}%
  \BibitemOpen
  \bibfield  {author} {\bibinfo {author} {\bibfnamefont {R.~A.}\ \bibnamefont
  {Waltz}}, \bibinfo {author} {\bibfnamefont {J.~L.}\ \bibnamefont {Morales}},
  \bibinfo {author} {\bibfnamefont {J.}~\bibnamefont {Nocedal}},\ and\ \bibinfo
  {author} {\bibfnamefont {D.}~\bibnamefont {Orban}},\ }\bibfield  {title}
  {\bibinfo {title} {An interior algorithm for nonlinear optimization that
  combines line search and trust region steps},\ }\href
  {https://doi.org/10.1007/s10107-004-0560-5} {\bibfield  {journal} {\bibinfo
  {journal} {Mathematical programming}\ }\textbf {\bibinfo {volume} {107}},\
  \bibinfo {pages} {391} (\bibinfo {year} {2006})}\BibitemShut {NoStop}%
\bibitem [{MAT()}]{MATLAB}%
  \BibitemOpen
  \href@noop {} {\bibinfo {title} {{MATLAB, version R2020a}}},\ \bibinfo {note}
  {{The MathWorks, Inc.}}\BibitemShut {Stop}%
\bibitem [{\citenamefont {Joubert}\ \emph {et~al.}(1987)\citenamefont
  {Joubert}, \citenamefont {Loisel}, \citenamefont {Chouan},\ and\
  \citenamefont {Haji}}]{joubert1987effect}%
  \BibitemOpen
  \bibfield  {author} {\bibinfo {author} {\bibfnamefont {P.}~\bibnamefont
  {Joubert}}, \bibinfo {author} {\bibfnamefont {B.}~\bibnamefont {Loisel}},
  \bibinfo {author} {\bibfnamefont {Y.}~\bibnamefont {Chouan}},\ and\ \bibinfo
  {author} {\bibfnamefont {L.}~\bibnamefont {Haji}},\ }\bibfield  {title}
  {\bibinfo {title} {The effect of low pressure on the structure of lpcvd
  polycrystalline silicon films},\ }\href@noop {} {\bibfield  {journal}
  {\bibinfo  {journal} {Journal of the Electrochemical Society}\ }\textbf
  {\bibinfo {volume} {134}},\ \bibinfo {pages} {2541} (\bibinfo {year}
  {1987})}\BibitemShut {NoStop}%
\bibitem [{\citenamefont {Voutsas}\ and\ \citenamefont
  {Hatalis}(1992)}]{voutsas1992structure}%
  \BibitemOpen
  \bibfield  {author} {\bibinfo {author} {\bibfnamefont {A.~T.}\ \bibnamefont
  {Voutsas}}\ and\ \bibinfo {author} {\bibfnamefont {M.~K.}\ \bibnamefont
  {Hatalis}},\ }\bibfield  {title} {\bibinfo {title} {Structure of as-deposited
  lpcvd silicon films at low deposition temperatures and pressures},\
  }\href@noop {} {\bibfield  {journal} {\bibinfo  {journal} {Journal of the
  Electrochemical Society}\ }\textbf {\bibinfo {volume} {139}},\ \bibinfo
  {pages} {2659} (\bibinfo {year} {1992})}\BibitemShut {NoStop}%
\bibitem [{\citenamefont {Yang}\ \emph {et~al.}(2000)\citenamefont {Yang},
  \citenamefont {Kahn}, \citenamefont {He}, \citenamefont {Phillips},\ and\
  \citenamefont {Heuer}}]{yang2000new}%
  \BibitemOpen
  \bibfield  {author} {\bibinfo {author} {\bibfnamefont {J.}~\bibnamefont
  {Yang}}, \bibinfo {author} {\bibfnamefont {H.}~\bibnamefont {Kahn}}, \bibinfo
  {author} {\bibfnamefont {A.-Q.}\ \bibnamefont {He}}, \bibinfo {author}
  {\bibfnamefont {S.~M.}\ \bibnamefont {Phillips}},\ and\ \bibinfo {author}
  {\bibfnamefont {A.~H.}\ \bibnamefont {Heuer}},\ }\bibfield  {title} {\bibinfo
  {title} {A new technique for producing large-area as-deposited zero-stress
  lpcvd polysilicon films: the multipoly process},\ }\href@noop {} {\bibfield
  {journal} {\bibinfo  {journal} {Journal of Microelectromechanical Systems}\
  }\textbf {\bibinfo {volume} {9}},\ \bibinfo {pages} {485} (\bibinfo {year}
  {2000})}\BibitemShut {NoStop}%
\end{thebibliography}%


%apsrev4-2.bst 2019-01-14 (MD) hand-edited version of apsrev4-1.bst
%Control: key (0)
%Control: author (8) initials jnrlst
%Control: editor formatted (1) identically to author
%Control: production of article title (0) allowed
%Control: page (0) single
%Control: year (1) truncated
%Control: production of eprint (0) enabled
\begin{thebibliography}{1}%
\makeatletter
\providecommand \@ifxundefined [1]{%
 \@ifx{#1\undefined}
}%
\providecommand \@ifnum [1]{%
 \ifnum #1\expandafter \@firstoftwo
 \else \expandafter \@secondoftwo
 \fi
}%
\providecommand \@ifx [1]{%
 \ifx #1\expandafter \@firstoftwo
 \else \expandafter \@secondoftwo
 \fi
}%
\providecommand \natexlab [1]{#1}%
\providecommand \enquote  [1]{``#1''}%
\providecommand \bibnamefont  [1]{#1}%
\providecommand \bibfnamefont [1]{#1}%
\providecommand \citenamefont [1]{#1}%
\providecommand \href@noop [0]{\@secondoftwo}%
\providecommand \href [0]{\begingroup \@sanitize@url \@href}%
\providecommand \@href[1]{\@@startlink{#1}\@@href}%
\providecommand \@@href[1]{\endgroup#1\@@endlink}%
\providecommand \@sanitize@url [0]{\catcode `\\12\catcode `\$12\catcode
  `\&12\catcode `\#12\catcode `\^12\catcode `\_12\catcode `\%12\relax}%
\providecommand \@@startlink[1]{}%
\providecommand \@@endlink[0]{}%
\providecommand \url  [0]{\begingroup\@sanitize@url \@url }%
\providecommand \@url [1]{\endgroup\@href {#1}{\urlprefix }}%
\providecommand \urlprefix  [0]{URL }%
\providecommand \Eprint [0]{\href }%
\providecommand \doibase [0]{https://doi.org/}%
\providecommand \selectlanguage [0]{\@gobble}%
\providecommand \bibinfo  [0]{\@secondoftwo}%
\providecommand \bibfield  [0]{\@secondoftwo}%
\providecommand \translation [1]{[#1]}%
\providecommand \BibitemOpen [0]{}%
\providecommand \bibitemStop [0]{}%
\providecommand \bibitemNoStop [0]{.\EOS\space}%
\providecommand \EOS [0]{\spacefactor3000\relax}%
\providecommand \BibitemShut  [1]{\csname bibitem#1\endcsname}%
\let\auto@bib@innerbib\@empty
%</preamble>
\bibitem [{\citenamefont {Wesemann}\ \emph {et~al.}(2019)\citenamefont
  {Wesemann}, \citenamefont {Panchenko}, \citenamefont {Singh}, \citenamefont
  {Della~Gaspera}, \citenamefont {G{\'o}mez}, \citenamefont {Davis},\ and\
  \citenamefont {Roberts}}]{WESEMANN2019SELECTIVE}%
  \BibitemOpen
  \bibfield  {author} {\bibinfo {author} {\bibfnamefont {L.}~\bibnamefont
  {Wesemann}}, \bibinfo {author} {\bibfnamefont {E.}~\bibnamefont {Panchenko}},
  \bibinfo {author} {\bibfnamefont {K.}~\bibnamefont {Singh}}, \bibinfo
  {author} {\bibfnamefont {E.}~\bibnamefont {Della~Gaspera}}, \bibinfo {author}
  {\bibfnamefont {D.~E.}\ \bibnamefont {G{\'o}mez}}, \bibinfo {author}
  {\bibfnamefont {T.~J.}\ \bibnamefont {Davis}},\ and\ \bibinfo {author}
  {\bibfnamefont {A.}~\bibnamefont {Roberts}},\ }\bibfield  {title} {\bibinfo
  {title} {Selective near-perfect absorbing mirror as a spatial frequency
  filter for optical image processing},\ }\href@noop {} {\bibfield  {journal}
  {\bibinfo  {journal} {APL Photonics}\ }\textbf {\bibinfo {volume} {4}},\
  \bibinfo {pages} {100801} (\bibinfo {year} {2019})}\BibitemShut {NoStop}%
\end{thebibliography}%
\end{document}

\preprint{APS/123-QED}

\title{Supplementary Material:\\High-NA optical edge detection via optimized multilayer films}

\author{Wenjin Xue}
\affiliation{Department of Electrical Engineering and Energy Sciences Institute, Yale University, New Haven, Connecticut 06511, USA}
\author{Owen D. Miller}
\affiliation{Department of Applied Physics and Energy Sciences Institute, Yale University, New Haven, Connecticut 06511, USA}

\date{\today}

\maketitle

\section{Detailed designs from Fig. 2(b,e) in main text}
\begin{figure}[htbp]
    \includegraphics[width=0.8\textwidth]{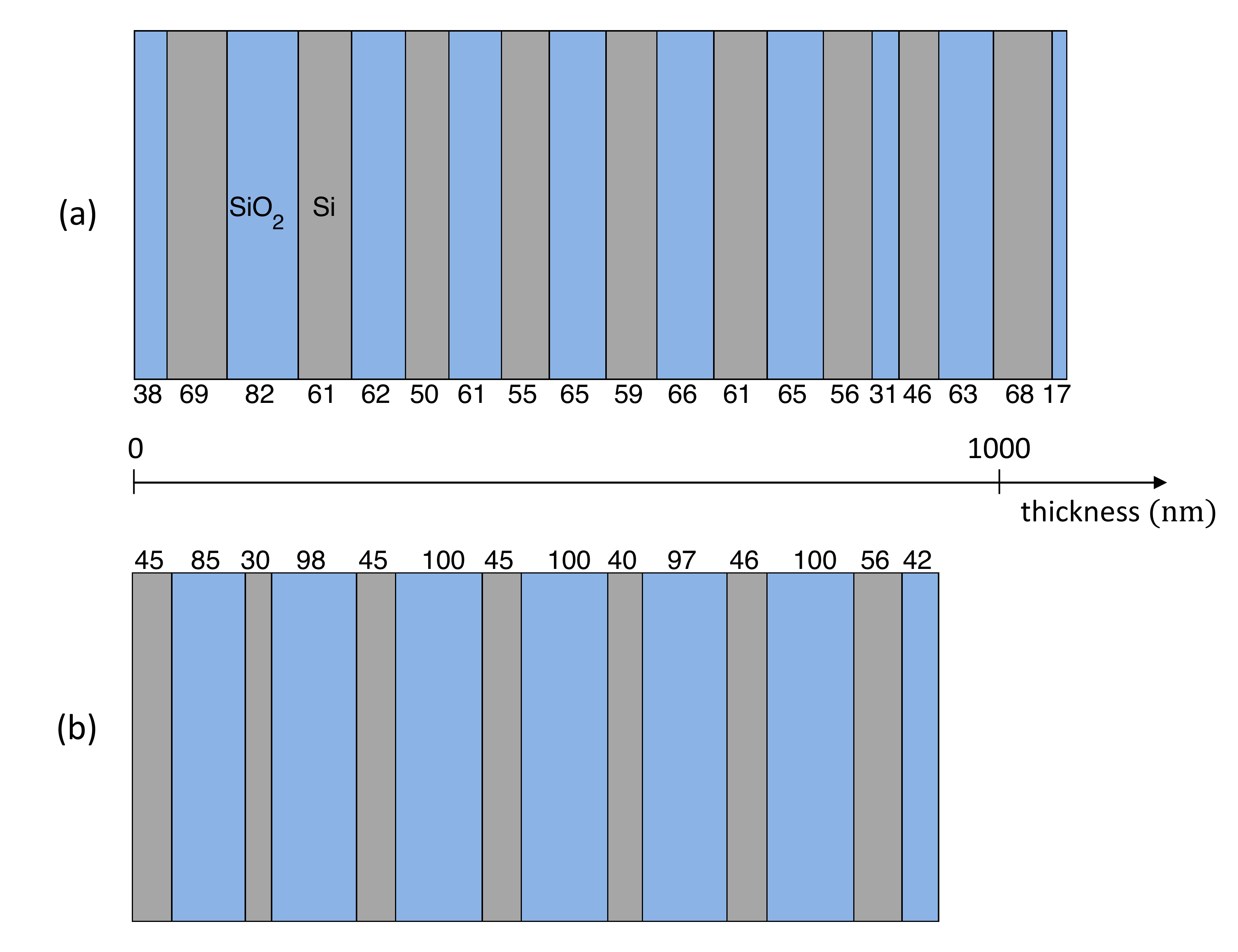}
    \centering
\caption{Detailed designs for Fig. 2(b,e) in main text, for (a) $t_{\rm target}\propto k_{\rho}^2$ and (b) $t_{\rm target}\propto \theta^3$.}
\label{fig:design_data}
\end{figure}

Here, in Figure~\ref{fig:design_data}, we provide the detailed thicknesses for the multilayer designs of Fig. 2(b,e) in the main text. The thicknesses of each layer are rounded to the nearest nanometer. The gray regions represent Si layers while the blue regions are SiO$_2$ layers. Both designs have a total thickness of roughly $\SI{1}{\um}$.

\section{Comparing the resolutions of our designs with those of \citeasnoun{WESEMANN2019SELECTIVE}}

\begin{figure}[htbp]
    \includegraphics[width=0.8\textwidth]{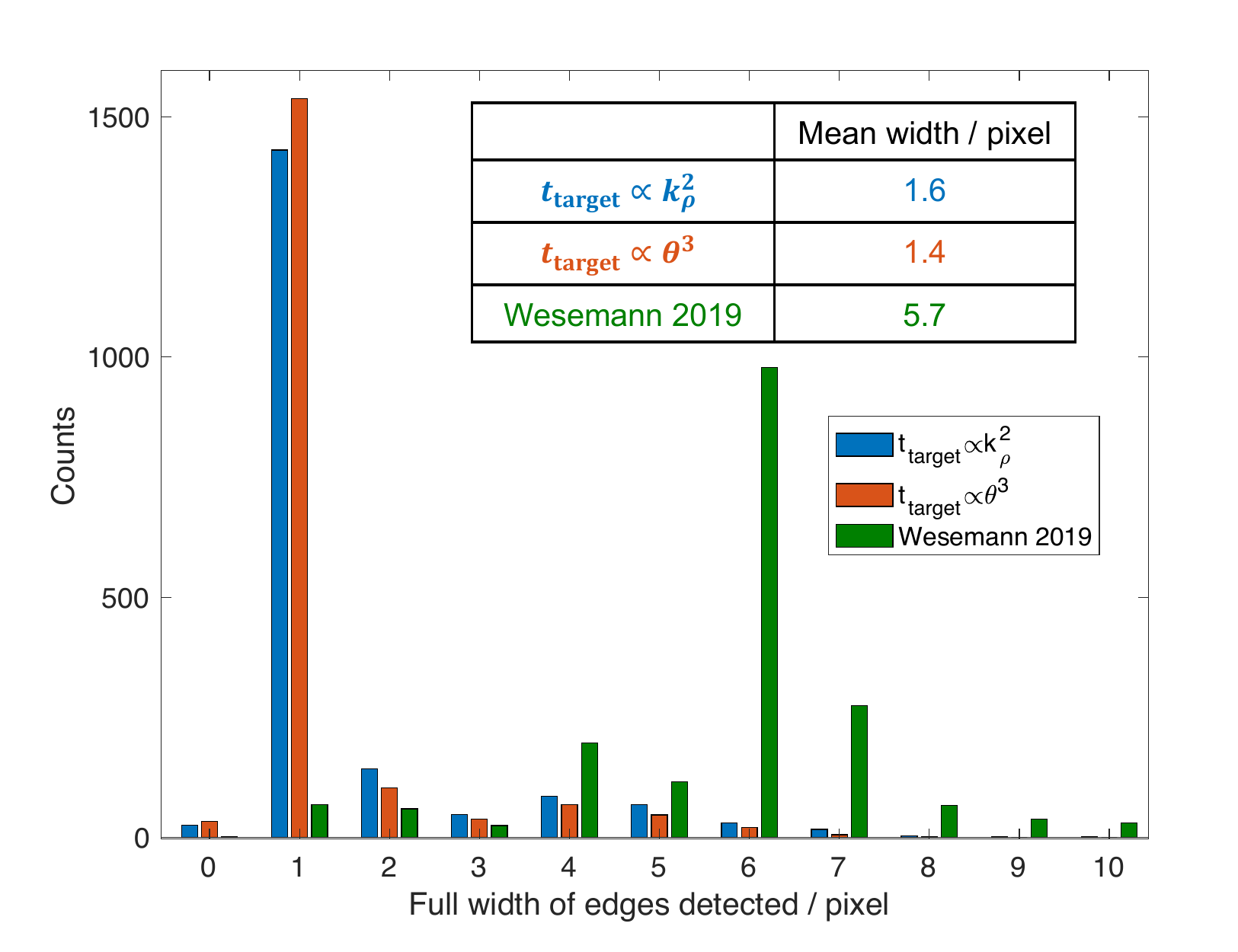}
    \centering
\caption{Width distributions of edges obtained from the plasmonic film design in \citeasnoun{WESEMANN2019SELECTIVE} (green bars) and multilayer designs in the main text (blue and red bars). The average width of the edges from the plasmonic film design in \citeasnoun{WESEMANN2019SELECTIVE} is 5.7 pixels, much larger than the average width from our multilayer designs.}
\label{fig:compare_edge_width}
\end{figure}

In Fig. 4 from the main text, the design structure denoted by a black cross, based on plasmonic films~\cite{WESEMANN2019SELECTIVE}, exhibits high-NA performance with ${\rm NA}=1$. However, this design works in reflection mode, requiring extra space for lenses and beam splitters. We also find that the phase variation of p-polarized reflection coefficients is a little more than \ang{180}, so the edges detected might lose significant clarity due to destructive interference. In Figure~\ref{fig:compare_edge_width}, we compare the resolutions of this design in reflection mode with our transmission-mode multilayer designs by comparing the widths of edges after simulation. In the simulation, the input figure is the same one from Fig. 2(a) in the main text, with the angular range of the input image take to correspond to ${\rm NA}=0.98$. The widths of the edges obtained from plasmonic film design in \citeasnoun{WESEMANN2019SELECTIVE} (green bars) have an average width of 5.7 pixels, and a modal edge width of 6 pixels, a few times larger than the output edges from our transmission-mode multilayer designs (blue and red bars).

\bibliography{SM_bib}